\documentclass{ws-procs9x6}

\begin{document}

\title{\large\lowercase{{\sc \uppercase{H}ermes} \uppercase{p}recision \uppercase{r}esults on $g_1^{\mathrm{p}}$, $g_1^{\mathrm{d}}$ and $g_1^{\mathrm{n}}$ and the \uppercase{f}irst \uppercase{m}easurement of the \uppercase{t}ensor \uppercase{s}tructure \uppercase{f}unction $b_1^{\mathrm{d}}$}}

\author{\lowercase{\uppercase{C}aroline \uppercase{R}iedl \emph{(on behalf of the {\sc \uppercase{H}ermes} collaboration)\\}}}

\address{Physikalisches Institut 2 der Universit\"at Erlangen-N\"urnberg, \\
Erwin-Rommel-Str. 1, D-91058 Erlangen, Germany\\ 
E-mail: criedl@mail.desy.de}

\maketitle

\abstracts{
{\bf Abstract.} Final {\sc Hermes} results on the proton, deuteron and neutron structure function $g_1$ are presented in the kinematic range $0.0021<x<0.9$ and $0.1<Q^2<20$ GeV$^2$. These results base on a refined analysis and are corrected for radiative and detector smearing effects using an unfolding algorithm. Furthermore, preliminary results on the first measurement of the tensor asymmetry $A_{zz}$ and the tensor structure function $b_1^{\mathrm{d}}$ are presented.}

\vspace{-0.25in}
Over the last 15 years, measurements of the polarized structure function $g_1$ by inclusive polarized deep-inelastic scattering (DIS)\cite{hepdata} have shed light on one basic ingredient of the spin puzzle: the quark contribution to the nucleon spin, $\Delta\Sigma$. For a spin-1 target, besides $g_1$ the quadrupole structure function $b_1$ is needed to parameterize the hadronic part of the interaction.\cite{b1theo} 

The final {\sc Hermes} $g_1$ results presented here base on a refined analysis of data taken with longitudinally polarized hydrogen\cite{g1p98} and deuterium\cite{lara} targets ($4$ resp.~$11.5\cdot 10^6$ DIS events), the $b_1$ results on a separate data set with a tensor-polarized deuterium target ($2.4\cdot 10^6$ DIS events).\cite{marco} The kinematic range covers $0.0021<x<0.9$ and $0.1<Q^2<20$ GeV$^2$. 

In the fixed-target experiment {\sc Hermes}, the 27.6 GeV lepton beam of the {\sc Hera} storage ring is scattered off an internal target of pure gases. The lepton beam is transversely self-polarized, for the reported data reaching an average polarization of $53\%$; longitudinal beam polarization in the interaction region is obtained by pairs of spin rotators. The {\sc Hermes} atomic gas target\cite{target} consists of an Atomic Beam Source (ABS) to generate and a Breit Rabi Polarimeter to measure nuclear polarization; a tubular open-ended, cooled storage cell is mounted inside the beam pipe confining the gas fed from the ABS. The average vector polarization was $85\%$, furthermore, the target design allowed for a simultaneous setting of an average tensor polarization of $-166\%$ and a residual vector polarization of only $+1\%$, which makes it possible to decouple the $b_1$ from the $g_1$ measurement.  

The {\sc Hermes} forward spectrometer\cite{spectrometer} includes numerous tracking chambers in front and behind a 1.3~Tm magnetic field and various detectors for particle identification. Electrons and hadrons are discriminated using a likelihood method based on the combination of four detectors.\cite{longdeltaq} The presented inclusive sample is contaminated less than $1\%$ with hadrons. 

The spin structure function $g_1(x,Q^2)$ is for both the proton and the deuteron determined from the ratio $g_1/F_1$ which is approximately equal to the virtual photon asymmetry $A_1$ measured via the longitudinal cross section asymmetry $A_{\|}$. For analysis details, see e.g.~Ref.~$[$\refcite{g1p98}$]$. For the neutron, $g_1^{\mathrm{n}}$ is calculated as a linear combination of $g_1^{\mathrm{p}}$ and $g_1^{\mathrm{d}}$.

\begin{figure}[ht]
\centerline{
\epsfxsize=1.8in\epsfbox[0 149 424 600]{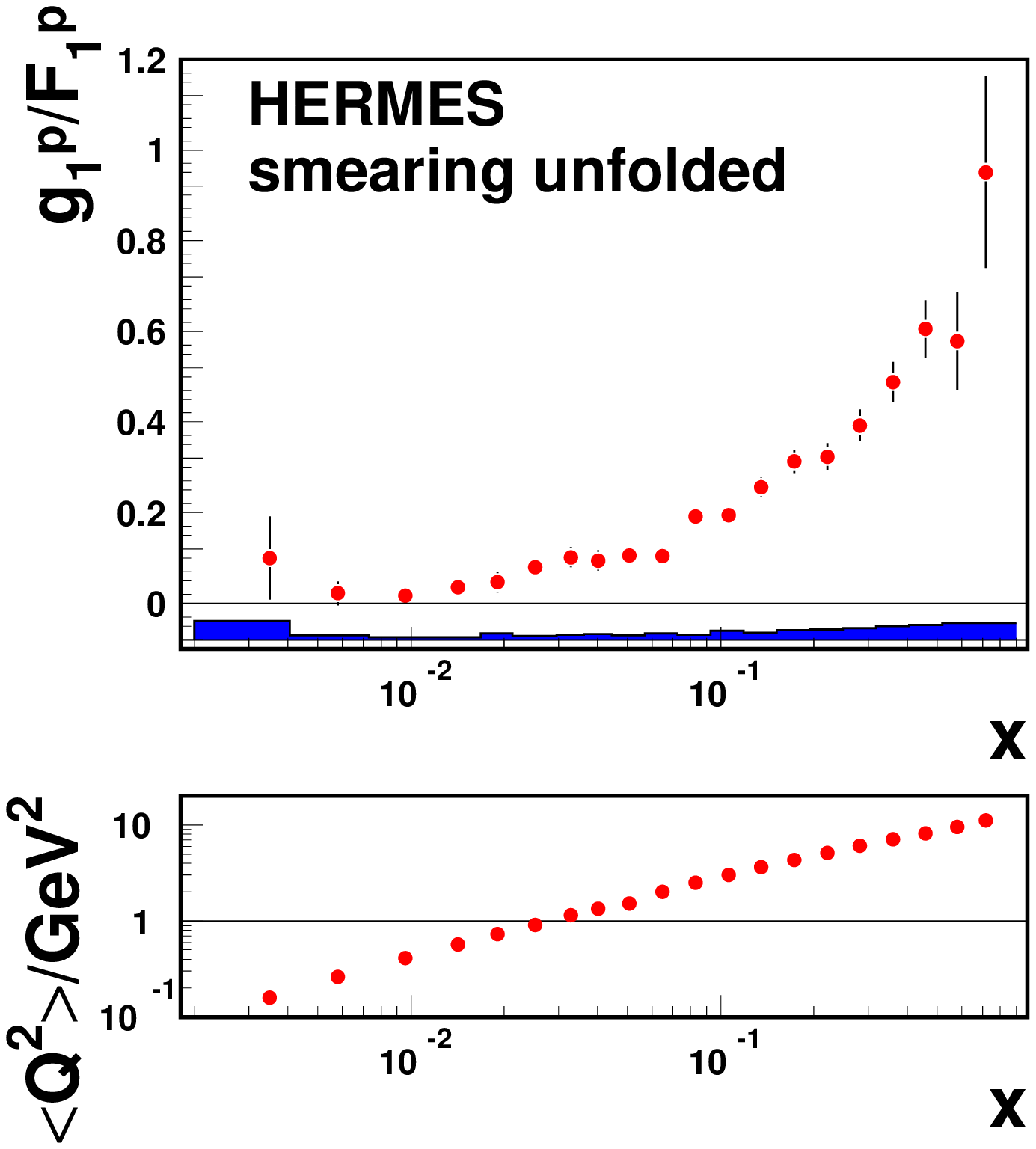}
\epsfxsize=1.8in\epsfbox[0 149 424 600]{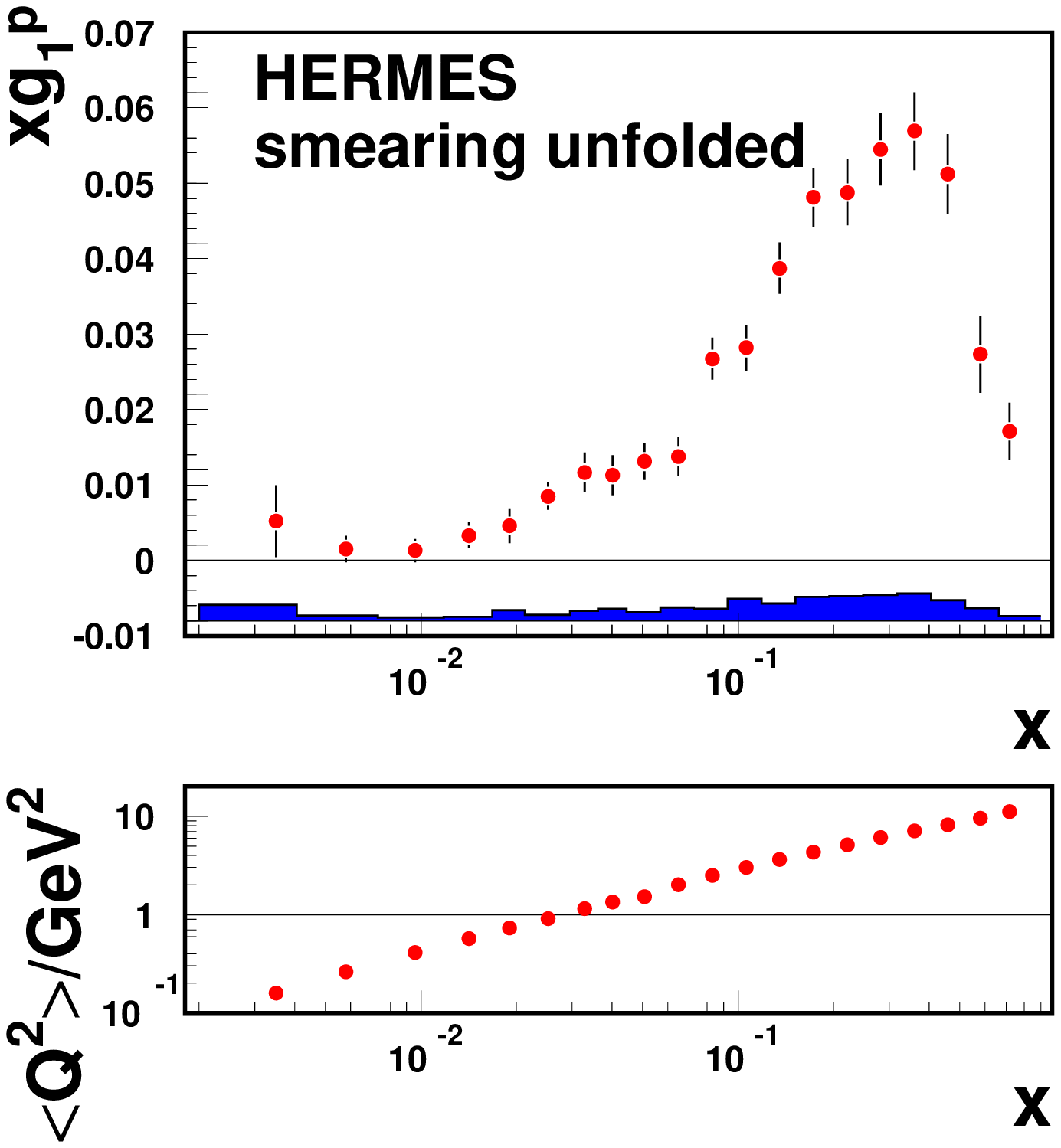}
\epsfxsize=1.5in\epsfbox[0 -150 568 568]{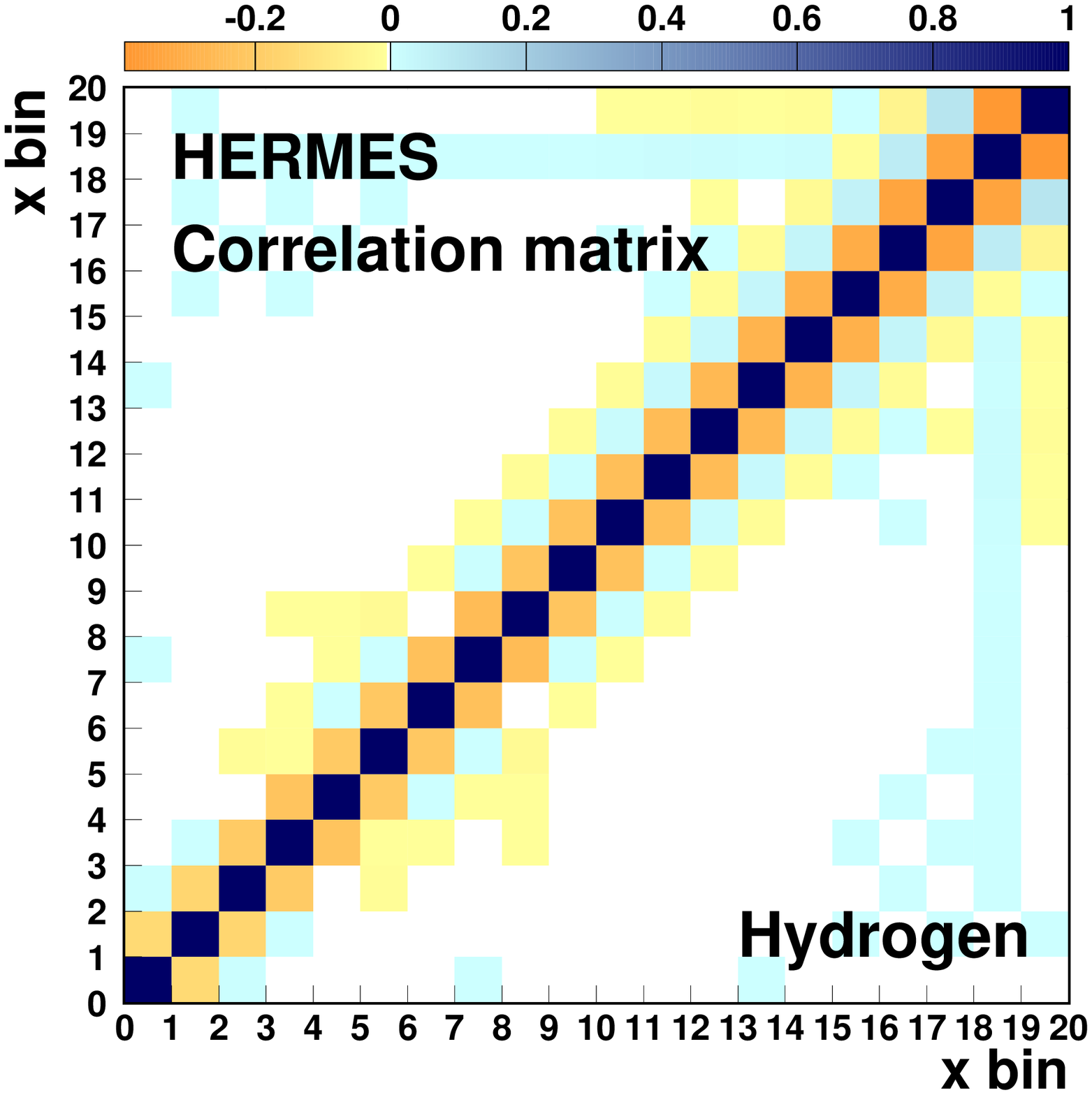}} 
\centerline{
\epsfxsize=1.8in\epsfbox[0 149 424 600]{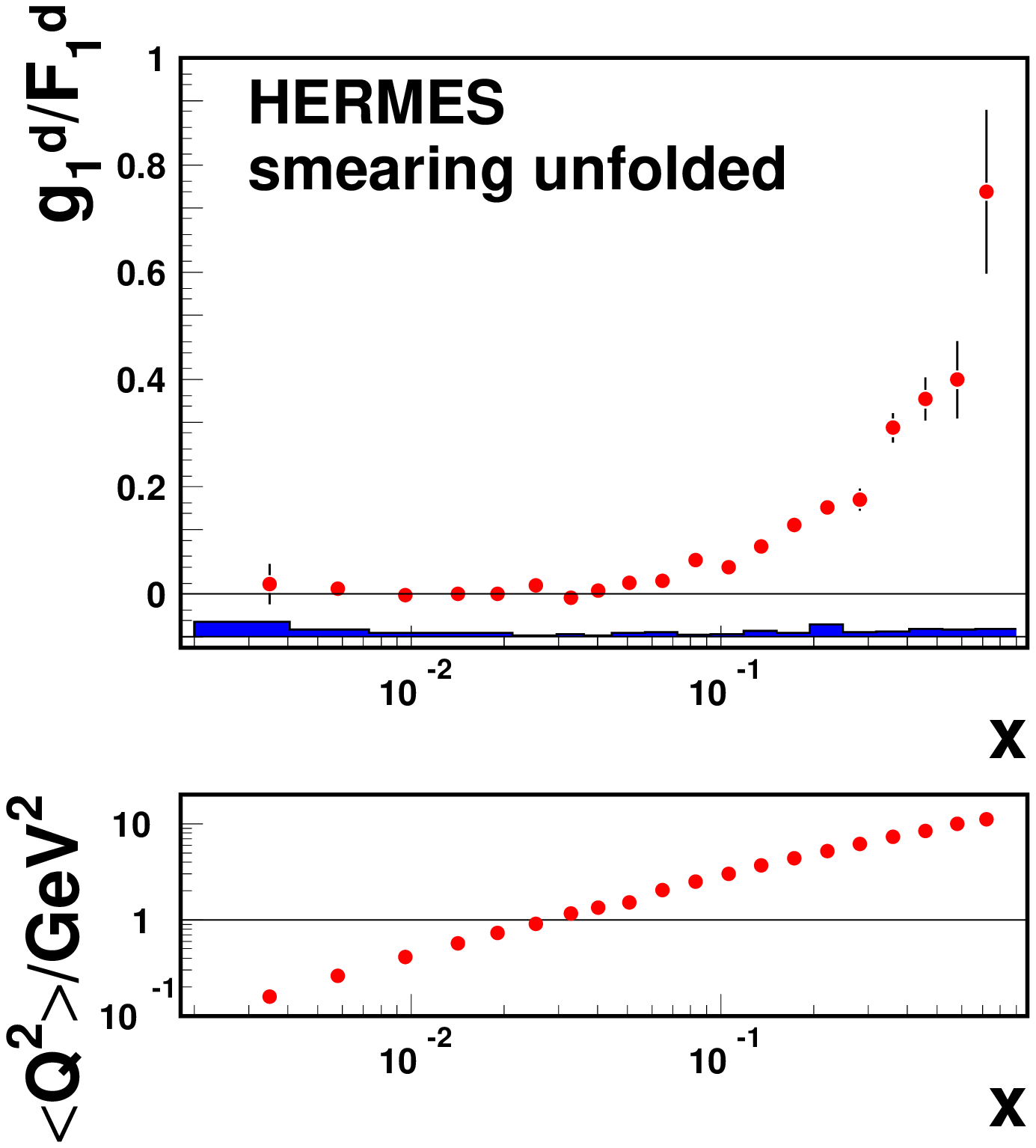}
\epsfxsize=1.8in\epsfbox[0 149 424 600]{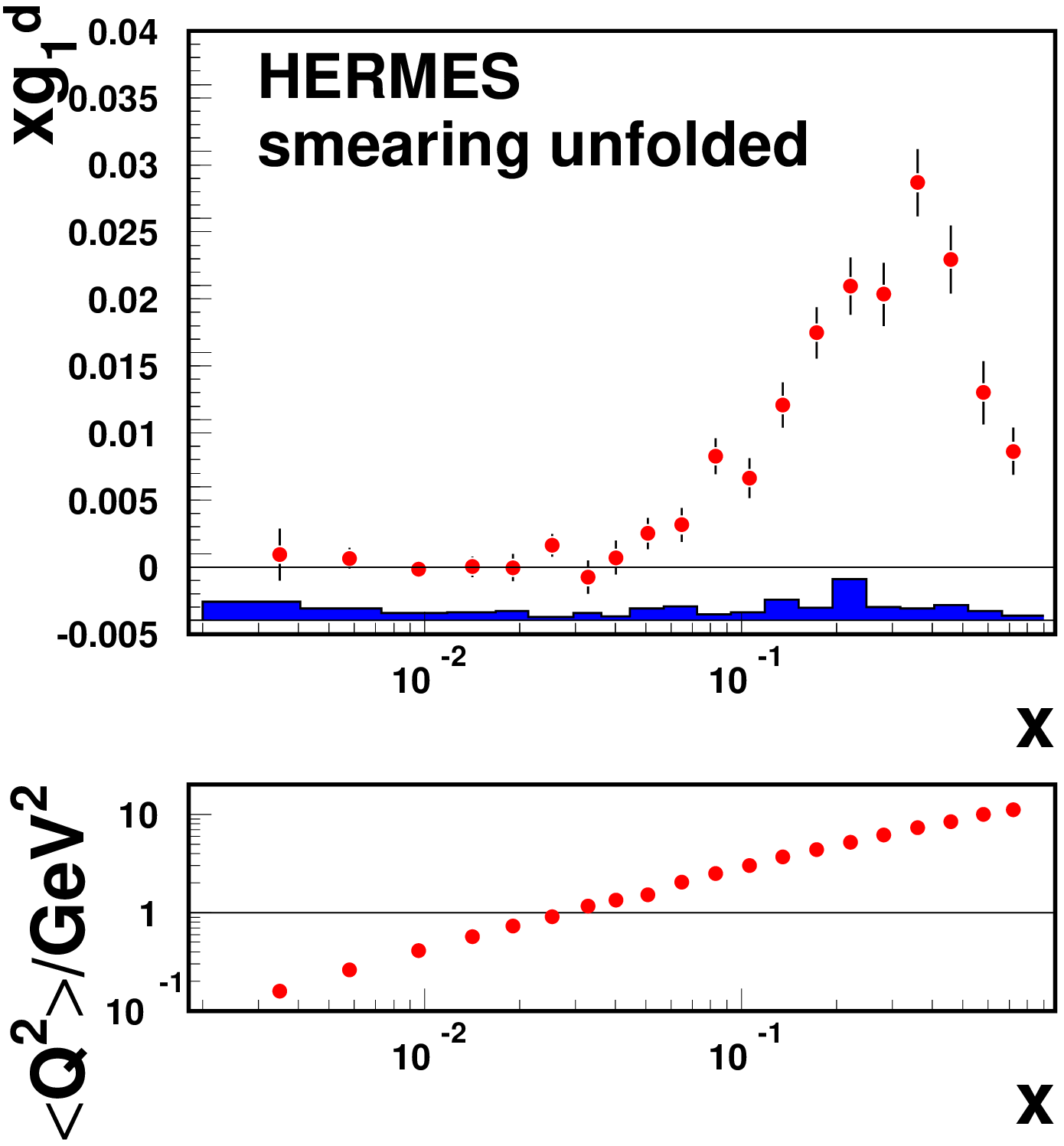}
\epsfxsize=1.5in\epsfbox[0 -150 568 568]{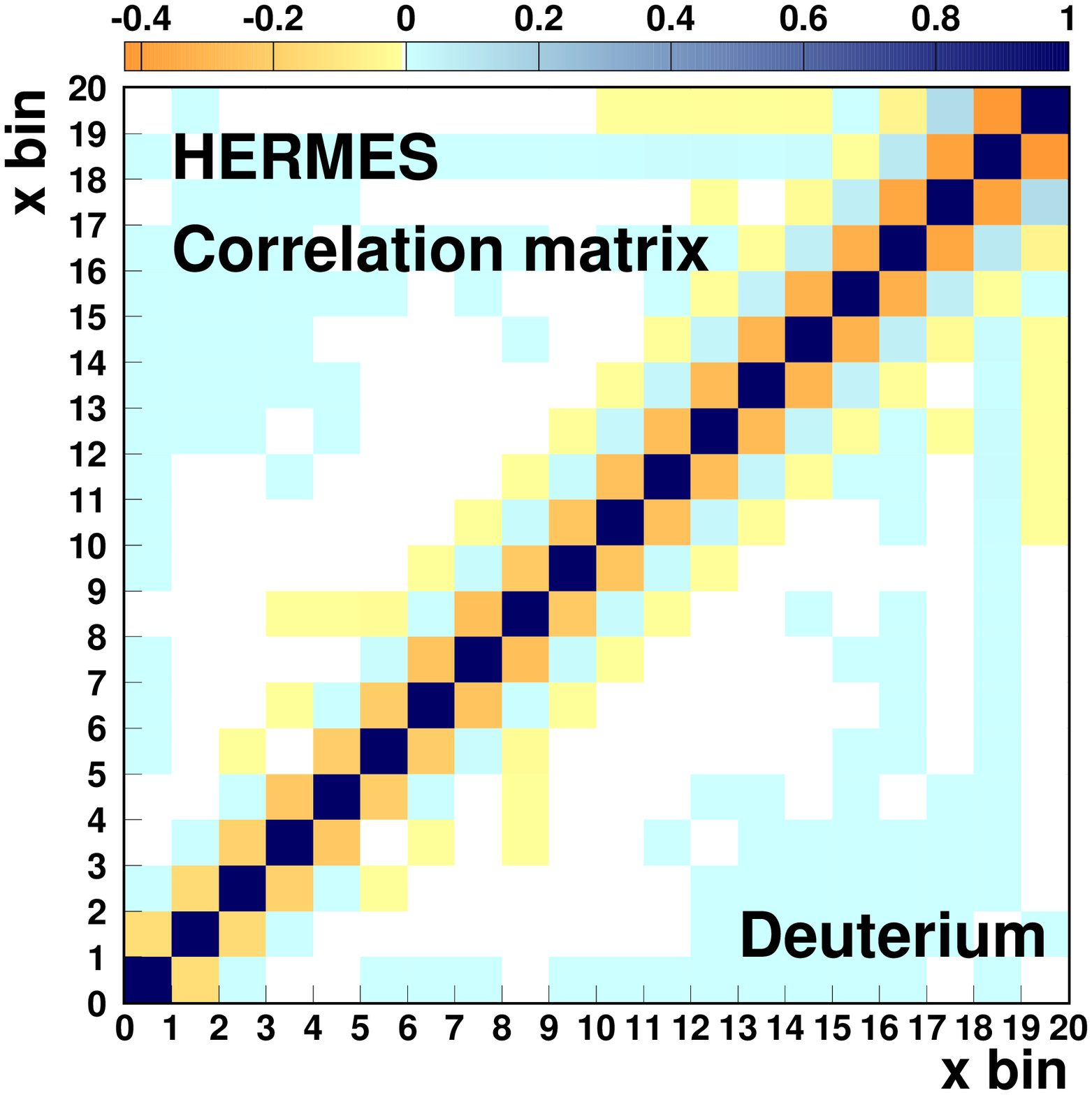}}    
\caption{\label{g1f1xg1} Smearing unfolded $g_1/F_1$ (left) and $xg_1$ (middle) as measured by {\sc Hermes} and the correlation matrix between statistical errors (right), for the proton (top row) and for the deuteron (bottom row). Error bars are statistical, displaying only the diagonal element of the correlation matrix, the shaded bands show the estimated systematic uncertainties}
\end{figure}

\begin{figure}[ht]
\centerline{
\epsfxsize=2.4in\epsfbox{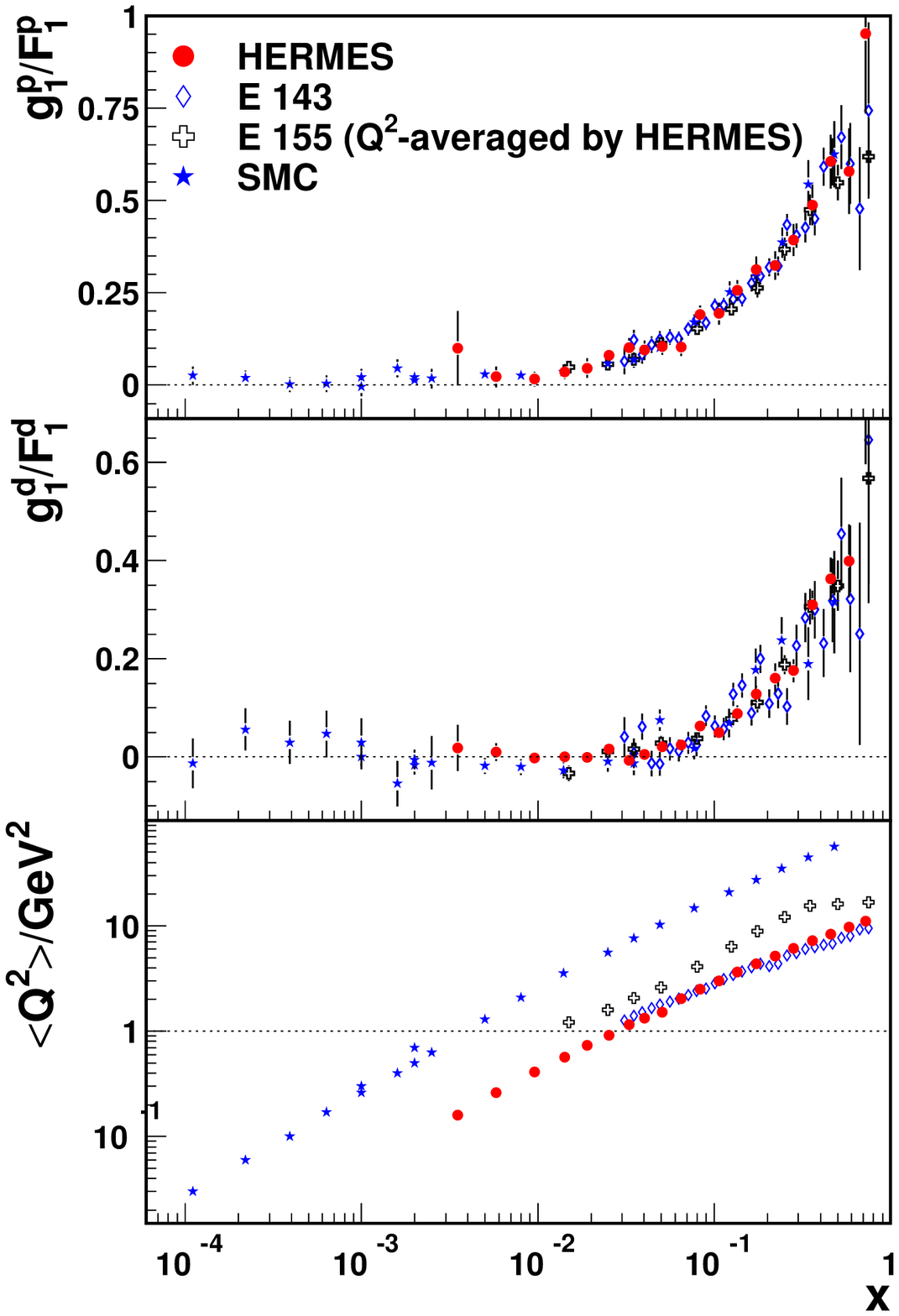}
\epsfxsize=2.4in\epsfbox{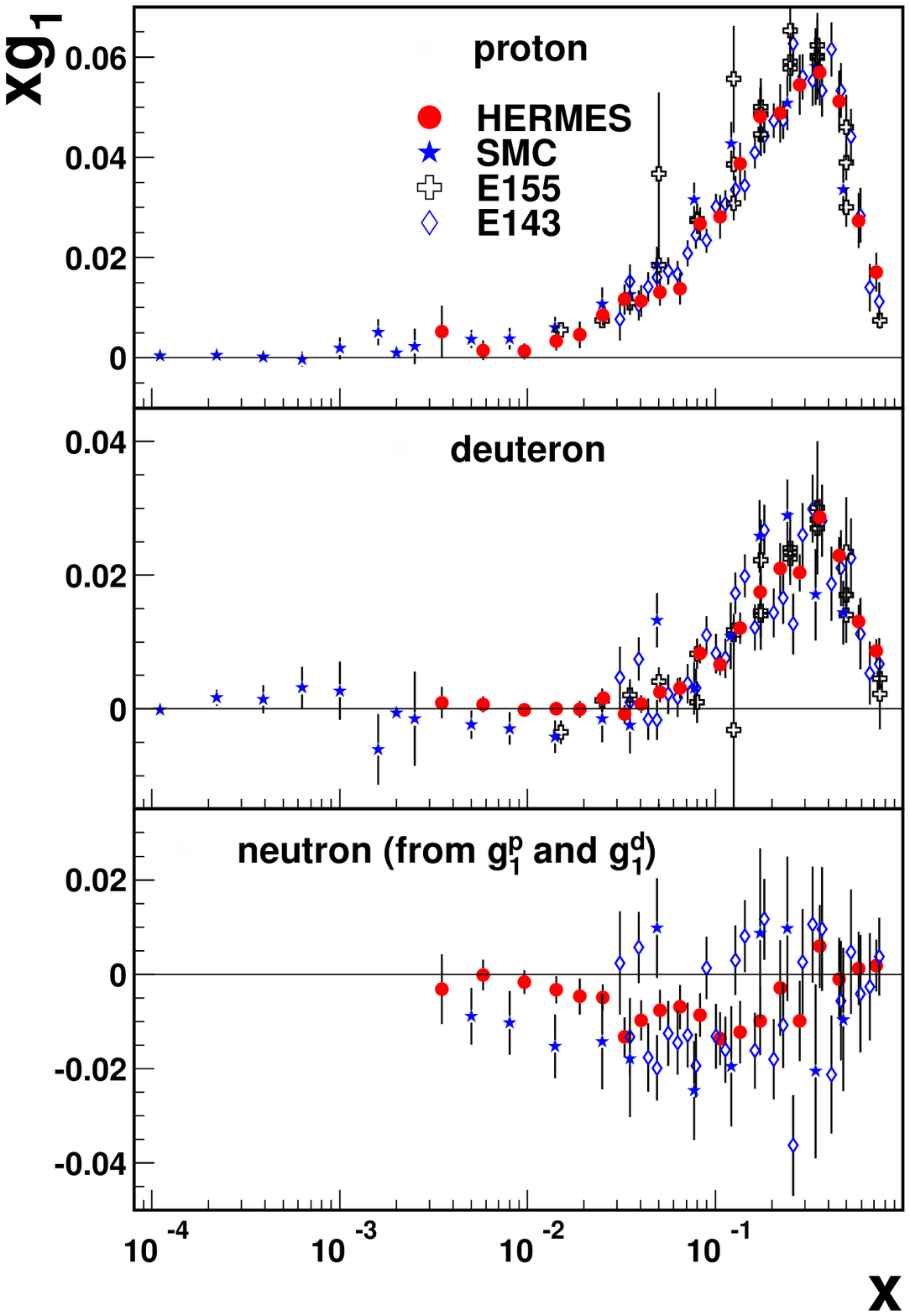}}  
\caption{\label{worlddata} {\sc Hermes} $g_1/F_1$ (left) and $xg_1$ data (right) in comparison to SLAC and CERN data for the proton (top panels) and the deuteron (middle panels) and $xg_1$ for the neutron from p and d data (right bottom panel) at the measured $Q^2$ (left bottom). Error bars show the quadratic sum of statistical and systematic uncertainties.}
\end{figure}

The measured asymmetries have been corrected for detector smearing and QED radiative effects to obtain the Born asymmetries.\cite{longdeltaq}$^,$\cite{g1long} These corrections have been applied using an unfolding algorithm that keeps track of the kinematic migration of events and doesn't require a fit on the data. The final Born asymmetries depend only on the measured data, on the detector model, on the known unpolarized cross sections and on the model for background processes. Figure \ref{g1f1xg1} shows the smearing unfolded $g_1/F_1$ and $xg_1$ results and,  based on the simulated event migration, their statistical correlations which must be taken into account when e.g.~calculating errors on moments or performing QCD fits to $g_1$ data.\cite{g1long} The applied algorithm removes systematic correlations between data points. A compilation of world data on $g_1/F_1$ and $xg_1$ is shown in Fig.~\ref{worlddata}.\cite{smclow}$^,$\cite{e143}$^,$\cite{e155} 

\begin{figure}[ht]
\centerline{
\epsfxsize=1.6in\epsfbox{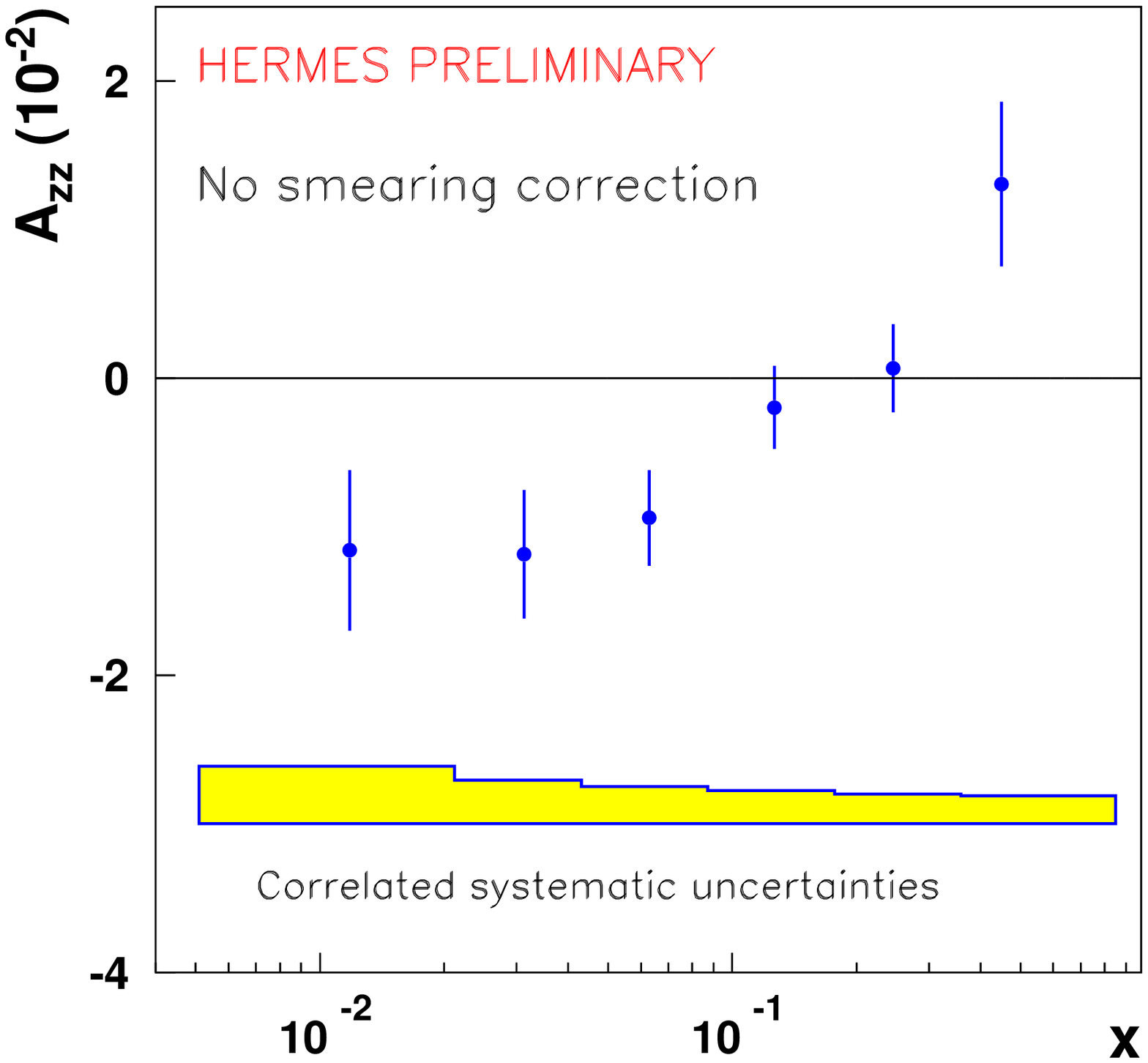}
\epsfxsize=1.4in\epsfbox[0 180 319 521]{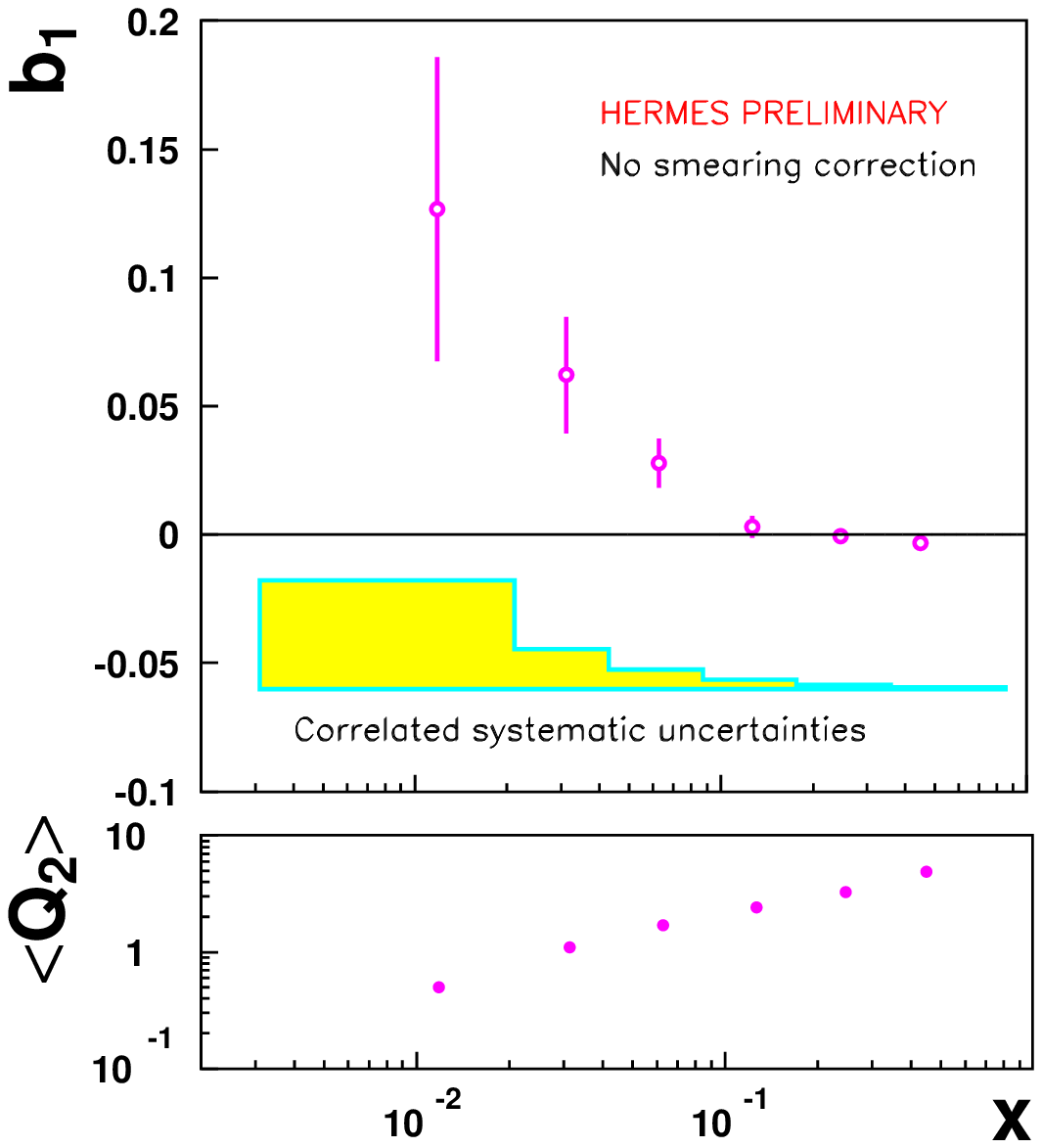}}   
\caption{\label{azzb1} The {\sc Hermes} tensor asymmetry $A_{zz}$ (left) and tensor structure function $b_1^{\mathrm{d}}$ (right; lower panel: average $Q^2$ for the measurements). The error bars are statistical only, the shaded bands show the estimated systematic uncertainties. }
\end{figure}

The tensor structure function $b_1(x,Q^2)$ is obtained from the ratio $b_1/F_1$, measured via $A_{zz}$ which compares the helicity-0 state of the deuteron with its averaged non-zero states. In Fig.~\ref{azzb1}, $A_{zz}$ and $b_1$ are displayed as measured for the first time by {\sc Hermes}; $A_{zz}$ is only of the order of $1\%$, implying the quadrupole contribution to the measurement of $g_1$ to be negligible. The steep rise of $b_1$ for small-$x$ suits latest model calculations.\cite{marco} The unfolding algorithm discussed above will also be applied to these data.\cite{b1} 

\section*{Acknowledgments}
I thank my {\sc Hermes} colleagues, especially the inclusive group. Furthermore, I gratefully acknowledge the financial support by the German Bundesministerium f\"ur Bildung und Forschung (contract number 06 ER 125I).


\end{document}